# Characteristics of Preferentially Attached Network Grown from Small World

**Seungyoung Lee**

*Graduate School of Innovation and Technology Management, Korea Advanced Institute of Science and Technology, Daejeon 305-701, Korea*

We introduce a model for a preferentially attached network which has grown from a small world network. Here, the average path length and the clustering coefficient are estimated, and the topological properties of modeled networks are compared as the initial conditions are changed. As a result, it is shown that the topological properties of the initial network remain even after the network growth. However, the vulnerability of each to preferentially attached nodes being added is not the same. It is found that the average path length rapidly decreases as the ratio of preferentially attached nodes increases and that the characteristics of the initial network can be easily disappeared. On the other hand, the clustering coefficient of the initial network slowly decreases with the ratio of preferentially attached nodes and its clustering characteristic remains much longer.

Email: sylee_inno@kaist.ac.kr

Fax: +82-42-350-4340

# I. INTRODUCTION

After the late 20 century, it has been found that a lot of real-world phenomena belong to complex network problems. Thus, complex networks have been studied in various topics such as the internet, social network, physics network, biological network, airline network, stock market and so on [1-8]. Numerous studies have discovered that complex networks in the real world share some common properties. Newman and Park found that a social network has a clustering characteristic, which means that a complex network has a tendency to have higher density of closed triads, which was originally proposed by Simmel [9, 10]. Clustering is widely observed in various networks from corporate board members to Hollywood movie actors [11, 12]. On the other hand, some real-world complex networks show a scale-free degree distribution which most nodes have only a few links, while some of them have extraordinarily many links [13]. Golder analyzed online friendship on Facebook and found that only a few users had more than 10,000 friends while the average number of Facebook friends is only 180 [14]. In this context, numerous complex network models have been studied to reveal the properties of complex systems. For instance, Erdös and Rényi suggested the random network model where isolated nodes are connected randomly with given probability [15]. Watt and Strogatz proposed the concept of small world network which is highly clustered and has small average path length and the algorithm to generate networks with the small world characteristics [16]. Barabási and Albert provided the evolving model to explain the power-law degree distribution and a simple way to generate scale-free networks [17].

In the real world, various complex networks show multiple properties at the same time. For example, scientific collaboration networks among researchers have clustering and preferential attachment characteristics. According to a study of scientific collaboration networks of more than 1.6 million researchers in physics and medicine, two researchers who have both cooperated with five common collaborators are about 200 times as likely to collaborate as a pair with none. Moreover, researchers who have already had a large number of collaborators are more likely to collaborate with other researchers in the future. It means that a degree distribution in scientific collaboration networks follows the power-law [18]. However, two representative complex network models, Watt and Strogatz model (WS model) and Barabási and Albert model (BA model), cannot represent clustering and scale-free characteristics simultaneously. The WS model can generate a complex network which has small-worldness and clustering properties, but the degree distribution is similar to that of a random graph. On the other hand, the BA model can generate a complex network which follows a power-law degree distribution, but its clustering coefficient is relatively low [13]. To resolve the problems, a lot of studies suggest alternative complex network models, which are simultaneously scale-free and small world [19-21]. Nonetheless, the WS model and the BA model have been widely used because of the algorithmic simplicity and the philosophical realization of the nature. Modifying these models is still an interesting subject [22].

Recent studies of social networks shows that the characteristics of an evolving network in the real world varies with time and that behavior of actors in a network can be changed as time passes by [23-25]. Moreover, these studies show behavioral differences between small and large group networks [26]. They show a possibility that some complex networks are constructed through multiple stages based on different behavior patterns. If a complex network evolves through multiple stages, the evolving result from the prior stage influences on the network structure of the next stage. Therefore, investigating a model which has separate stages based on the different algorithm can give insights into real world networks.

In this manner, we suggest the modified BA model to describe an evolving process divided in two parts. We propose a model for generating a preferentially attached network which varies with the initial network structure and show topological properties of these networks. We mainly estimate the average path length and the clustering coefficient. These parameters represent a network structure, so they show an influence of the initial condition on topological properties of the preferentially attached network. In section II, the detail of the model is introduced. Numerical calculations are performed and the result is shown in Section III. Conclusion is in Section IV.

## II. MODEL AND METHOD

We introduce a model for generating graph which has grown from small world networks by using the preferential attachment. In the standard BA model, we usually start with a small number of nodes and add new nodes and edges to the network at every time step. This algorithm well represents the open systems and growing process of the real world networks by continuously adding new nodes. However, some growing complex networks in the real world may not follow the consistence growing algorithm in evolution process [27]. According to a study on collaborations between scientists, small groups of scientists rely on few strong connections, while large groups show a high turnover of the members [26]. It means that the behavior of actors in a network may depend on the network's scale or stage. In this context, we suggest a model with two stages which generates preferentially attached networks, which start with a network already constructed. We use small world networks with different size and $\beta$ as an initial condition and grow these networks based on the preferential attachment.

The model has two steps, generating a small world network based on the WS model and growing a network based on the BA model. The definition of the model is as follows. Step1. Generating a small world. First of all, we generate a ring lattice with $N_1$ vertices, each of which is connected to its closest $K_1$ neighbor. Every edge is undirected and unweighted. Then, we choose a vertex and reconnect its clockwise closest edge to another random non-linked vertex with probability $\beta$, and repeat this process by moving clockwise around the ring. After one lap, we repeat this rewiring process with the next clockwise closest edges until every edge of the original lattice is considered once [16]. Step2. Growing the network. First of all, we add a new vertex to the network from Step1. Then, we add $K_2$ edges which connect $K_2$ different vertices in the network with the new vertex. We assume that the new vertex would follow the preferential attachment and the probability of connecting to vertex i is depend on the connectivity $k_i$ of that vertex.

$$p_i = \frac{k_i}{\sum_j k_j}$$

We repeat this until $N_2$ vertices are added to the initial network [17].

The aim of this study is to see how the structures of preferentially attached networks vary with different initial structure. We compare the network structures by estimating the average path length and the clustering coefficient. The average path length is a concept which represents the average number of the steps it takes to get from one vertex to another. It is defined as

$$\ell = \frac{1}{N(N-1)} \sum_{i \neq j} d(v_i, v_j)$$

, where N is the total number of vertices and $d(v_i, v_j)$ is the shortest path between vertex i and j [13]. The clustering coefficient is a measure of the degree in which vertices clustered together. There are two types of clustering coefficient exist, the global and the local. The global clustering coefficient is a fraction of closed triplets to all triplets in the whole network. It is defined as

$$C_g = \frac{Number\ of\ closed\ triplets}{Number\ of\ connected\ triplets\ of\ vertices}$$

This gives information of clustering in a whole complex network [28]. The local clustering coefficient shows how close the vertex's neighbors are to a complete graph. The local clustering coefficient of vertex i is defined as

$$C_i = \frac{2 E_i}{k_i(k_i - 1)}$$

, where $k_i$ is the number of links of the vertex i and $E_i$ is the number of links that exist among its nearest neighbors. As an alternative to the global clustering coefficient, the average of the local clustering coefficients of all vertices is used as a measure of clustering in the whole network [16].

$$C_l = \frac{1}{N} \sum_i^N C_i$$

We mainly use the average local clustering coefficient as a measure of clustering in this study.

In the following simulations, we set $N = N_1 + N_2 = 1000$ and $K_1 = 10$.

### III. SIMULATION RESULTS

At first, we set $K_2 = 5$ to maintain the total degree regardless of change in $r = \frac{N_2}{N}$. In the case of $r = 1$, this model is same as the BA model, and in the case of $r = 0$, this model is same as the WS model. In figure 1, we compare the average path length $\ell$ and the clustering coefficient by changing the rewiring probability β and the ratio of preferentially attached node r.

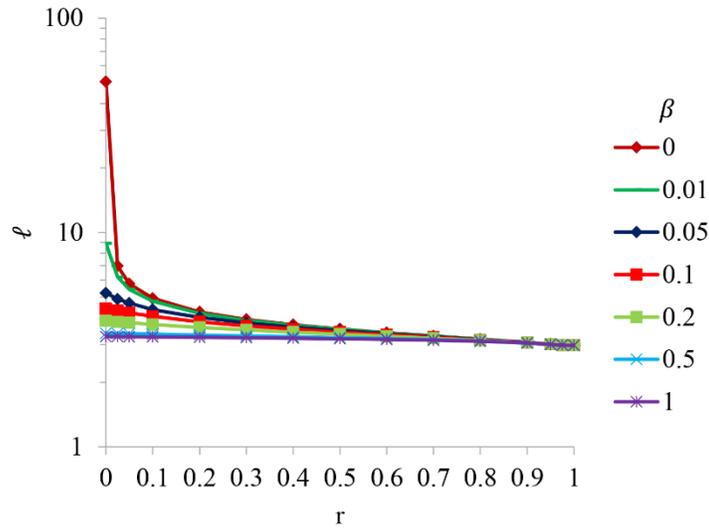

**Fig. 1. Plot of the average path length verses the ratio of preferentially attached nodes r. And all observables are from averages over 100 network realizations.**

In figure 1, we measure the average path length of the modeled networks. The result shows that the average path length decreases with r in every case of β. The average path length rapidly declines even when a small number of preferentially attached nodes are added to the small world network. This effect becomes bigger when β is smaller. This means that preferentially attached nodes act like bridges between distant nodes. If β is small, the initial network is similar to a ring lattice, so most of the initial nodes are only linked with their neighbors. On the contrary, a newly added node makes $K_2$ links with the vertices selected from the whole network. Thus, it is possible that a newly added node connects the distant nodes indirectly. Moreover, as more nodes are added, the possibility of connecting a distant node to hub nodes increases because of the preferential attachment. Therefore, adding preferentially attached nodes can dramatically reduce the average distant between nodes with a few nodes.

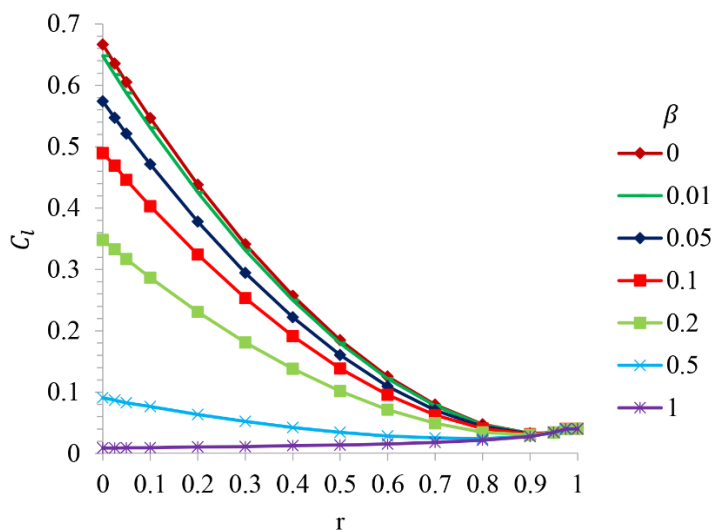

**Fig. 2. Plot of the average local clustering coefficient verses the ratio of preferentially attached nodes r.**

Figure 2 shows the average local clustering coefficient $C_l$ of the modeled networks. Like the average path length, the average local clustering coefficient converges on that of the BA model as r increases. Therefore, it decreases with r when β is small and increases when β is large. However, its convergence is slower than the average path length's. It indicates that adding preferentially attached nodes reduces clustering because they build links with the distant nodes which are not linked with each other when β is small, but its impact is limited in the nodes connected directly. So, the clustering coefficient decreases with the accumulation of preferentially attached nodes. On the other hand, when β is large, adding preferentially attached nodes increases clustering, because it develops hub nodes. A network with hub nodes shows higher clustering than a random network because hub nodes increase the probability of generating triplet. As a result, the clustering coefficient increases with the accumulation of preferentially attached nodes. We also calculate the global clustering coefficient and see the same pattern with the average of local clustering coefficient. One more interest thing is that the averaged local clustering coefficient increases with the ratio of preferentially attached nodes when the ratio is very high. It indicates that the modeled network can be less clustered than a fully preferentially attached network within the specific range of r even when it starts with highly clustered one.

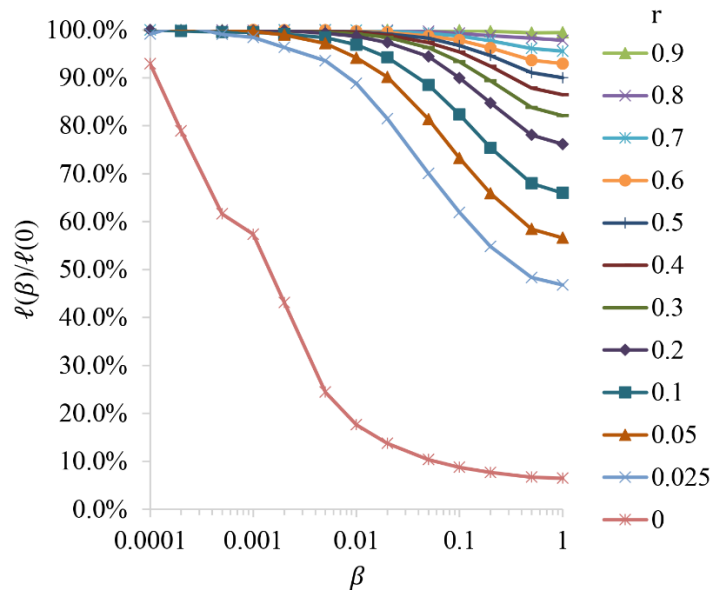

**Fig. 3. Plot of the normalized average path length verses the rewiring probability of initial network β.**

To investigate the influence of initial network's β, we normalize the average path length and the clustering coefficient by those in the case of $β = 0$. Figure 3 shows the normalized average path length. The normalized average path length decreases with β in every case of r and this is similar to the WS model's characteristic. However, the influence of β dramatically decreases with r. We can see that the relative size of the average path length between β = 0.0001 and β = 0 decrease to about 50% when only 2.5% of preferentially attached nodes are added. It indicates that the initial network's average path length influences on the average path length of the final network, but its influence is vulnerable to preferentially attached nodes being added.

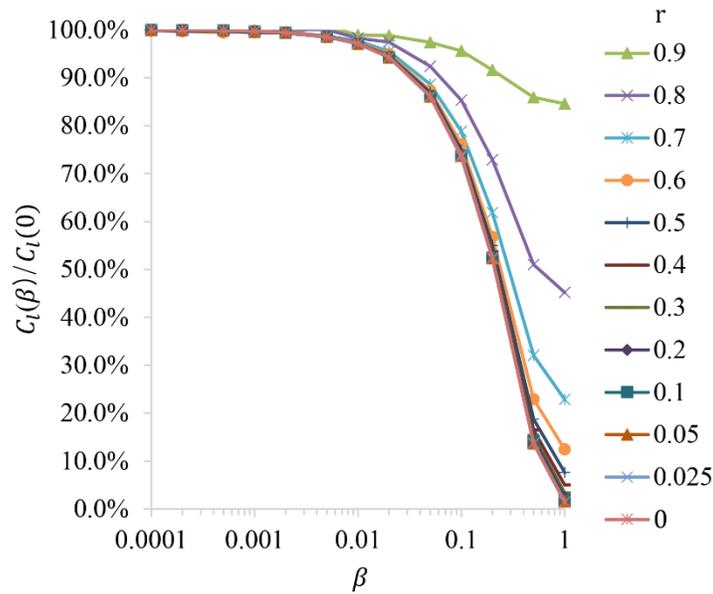

**Fig. 4. Plot of the normalized average local clustering coefficient verses the rewiring probability of initial network β.**

In figure 4, we show the normalized average local clustering coefficient. It also shows that the normalized average local clustering coefficient decreases with β, and its pattern is similar to the WS model's. However, in contrast to the average path length, the influence of β is not sensitive to the changes in r. The result of each case show similar normalized curves through β until the preferentially attached nodes occupy more than half of the network. It indicates that the initial network's clustering property remains even after the network's growth.

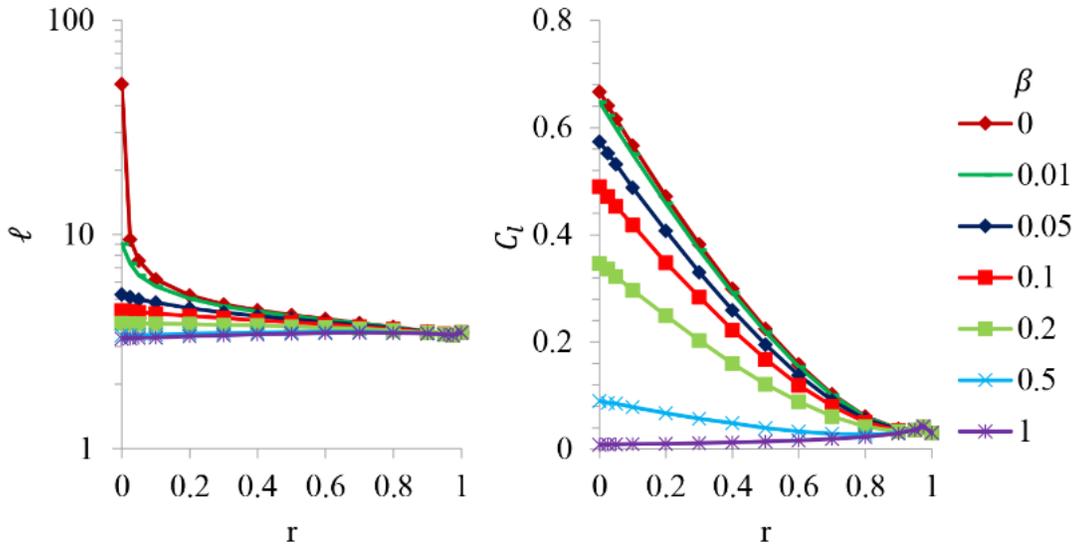

**Fig. 5. Plots of the average path length and the clustering coefficient of the case when $K_2 = 3$ verses the ratio of preferentially attached nodes r.**

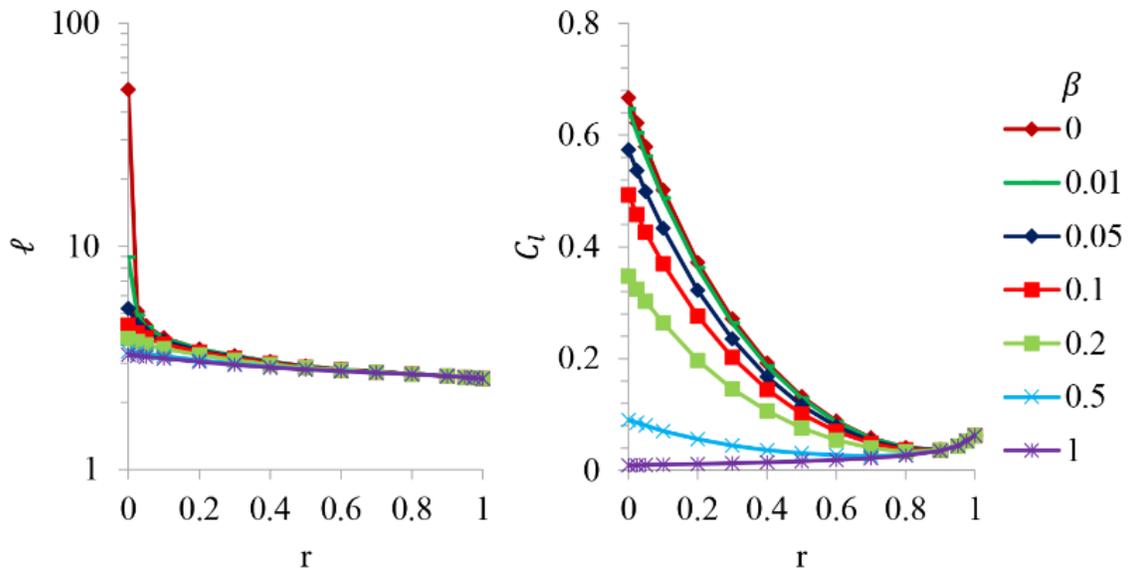

**Fig. 6. Plots of the average path length and the clustering coefficient of the case when $K_2 = 10$ verses the ratio of preferentially attached nodes r.**

We also investigate the case of $K_2 = 3$ and 10 to check the robustness. Figure 5 shows the result of the case when $K_2 = 3$, where the total degree decreases with r, and figure 6 shows the result of the case when $K_2 = 10$, where the total degree increases with r. In both cases, we can see that the average path length and the average local clustering coefficient show the similar results to that with $K_2 = 5$.

## IV. CONCLUSION

In this paper, we suggest the modified BA model to describe an evolving process divided into two stages. We use a rewired ring lattice as a seed network and evolve it using preferential attachment. We then investigate how the structural properties of modeled network vary with structures of the initial network. As a result, we show that topological properties of the initial network remain even after the network growth. However, its vulnerability to preferentially attached nodes being added is not the same. We show that the average path length rapidly decreases with the ratio of preferentially attached nodes, which is because the nodes that are newly added act like bridges, and the characteristics of the initial networks can be removed easily. On the other hand, the clustering coefficient of the initial network slowly decreases when ratio of preferentially attached nodes increase, which is because the influence of a new node is limited in the nodes connected directly, and its clustering characteristic remains much longer. Because of these differences, the modeled network with low ratio of preferentially attached nodes and small β shows small world properties, the high clustering coefficient, and the small average path length. We hope that our findings help increase understanding of the preferential attachment process and its characteristics.

## REFERENCES

[1] M. E. J. Newman, Networks: An Introduction (Oxford University University Press, Oxford, 2010)

[2] S. -H. Yook, H. Jeong and A. -L. Barabasi, Proc. Nat. Acad. Sci. 99, 13382 (2002)

[3] Y. Kim, J. -H. Kim and S. -H. Yook, Phys. Rev. E 83, 056115 (2011)

[4] S. Min and K. Kim, J. Korean Phys. Soc. 65, 1164 (2014)

[5] H. -J. Kim, I. -M. Kim, Y. Lee and B. Kahng, J. Korean Phys. Soc. 40, 1105 (2002)

[6] S. -H. Yook, Z. Oltvai and A. -L. Barabasi, Proteomics 4, 928 (2003)

[7] R. Guimerà and L. A. N. Amaral, Eur. Phys. J. B 38, 381 (2004)

[8] H. Ebel, L. -I. Mielsch and S. Bornholdt, Phys. Rev. E 66, 035103(R) (2002).

[9] M. E. J. Newman and J. Park, Phys. Rev. E 68, 036122 (2003)

[10] G. Simmel, The Sociology of Georg Simmel. (Free Press, New York, 1950)

[11] G. Davis, M. Yoo and W. E. Baker, Strateg. Organ. 3, 301 (2003)

[12] D. J. Watts, Am. J. Sociol. 105, 493 (1999)

[13] R. Albert and A. -L. Barabàsi, Ref. Mod. Phys. 74, 47 (2002)

[14] S. Golder, D. Wilkinson and B. Huberman, 3rd International Conference on Communities and Technologies (CT2007), East Lansing, MI. June 28 (2007)

[15] P. Erdös and A. Rényi, Publicationes Mathematicae 6, 290 (1959)

[16] D. J. Watts and S. H. Strogatz, Nature 393, 440 (1998)

[17] A. -L. Barabàsi and R. Albert, Science 286, 509 (1999)

[18] M. E. J. Newman, Phys. Rev. E 64, 025102R (2001)

[19] K. Klemm and V. M. Eguíluz, Phys. Rev. E 65, 057102 (2002)

[20] F. Comellas, G. Fertin and A. Raspaud, Phys. Rev. E 69, 037104 (2004)

[21] J. S. Andrade Jr., H. J. Herrmann, R. F. S. Andrade and L. R. da Silva, PRL 94, 018702 (2005)

[22] X. Li and G. Chen, Physica A 328, 274 (2003)

[23] J. Leskovec, J. Kleinberg and C. Faloutsos, in Proceedings of the eleventh ACM SIGKDD international conference on Knowledge discovery in data mining. (ACM, 2005) p. 177-187

[24] P. Holme, C. R. Edling and F. Liljeros, Social Networks 26, 155 (2004)

[25] R. Kumar, J. Novak, P. Raghavan and A. Tomkins, World Wide Web 8, 159 (2005)

[26] G. Palla, A. -L. Barabási and V. Tamás, Nature 446, 664 (2007)

[27] G. Paperin, D. G. Green, S. Sadedin, J R Soc Interface 8, 609 (2011)

[28] R. D. Luce and A. D. Perry, Psychometrika 14, 95 (1949)